\documentclass[dvips]{article}
\usepackage[centertags]{amsmath}
\usepackage{amssymb,amscd}
\usepackage{icrctc07}
\newcommand{\cA}{\phantom{0}}
\title{
Sensitivity studies for the cubic-kilometre deep-sea neutrino telescope KM3NeT
}
\shorttitle{ Sensitivity studies for KM3NeT }
\authors{ J. Carr$^{1}$, D. Dornic$^{1}$, F. Jouvenot$^{2}$, U.F. Katz$^{3}$, S. Kuch$^{3}$, 
        G. Maurin$^{4}$, R. Shanidze$^{3}$,\\ on behalf of the KM3NeT consortium}
\shortauthors{J. Carr, D. Dornic, F. Jouvenot, U.F. Katz, S. Kuch, G. Maurin, R. Shanidze,
              representing the KM3NeT consortium}
\afiliations{
$^1$ CPPM - Centre de Physiques des Particules de Marseille, CNRS/IN2P3 et Universite de
Mediterranee, 163 avenue de Luminy, Case 902, 13288 Marseille Cedex 9, France\\
$^2$ University of Liverpool, Oliver Lodge Laboratory, L69 7ZE, Liverpool, United Kingdom\\
$^3$ University Erlangen-Nuremberg, Physics Institute, Erwin-Rommel-Str.\ 1, Erlangen, 91058,
     Germany\\
$^4$ DSM/DAPNIA - Direction des Sciences de la Matiere, laboratoire de recherche sur les lois
     fondamentales de l'Univers, CEA Saclay, 91191 Gif-Sur-Yvette Cedex, France}

\abstract
{
The observation of high-energy neutrinos from astrophysical sources would
substantially improve our knowledge and understanding of the non-thermal
processes in these sources, and would in particular pinpoint the accelerators of
cosmic rays. The sensitivity of different design options for a future
cubic-kilometre scale neutrino telescope in the Mediterranean Sea is
investigated for generic point sources and in particular for some of the
galactic objects from which TeV gamma emmission has recently been observed by
the H.E.S.S.\ atmospheric Cherenkov telescope. The effect of atmospheric
background on the source detection probabilities has been taken into account
through full simulation. The estimated event rates are compared to previous
results and limits from present neutrino telescopes.
}

\begin{document}
\maketitle

\section{Introduction}

The KM3NeT consortium {\cite{KM3NeT-1}} is currently working on a conceptual
design for a future Mediterranean neutrino telescope, which will have an
instrumented volume of a scale of one km$^3$.
 
One of the goals of the KM3NeT Design Study is the optimisation of the detector
configuration with respect to the observation of astrophysical neutrino sources. 
The observation of high-energy neutrino point sources will substantially improve
our knowledge and understanding of the non-thermal processes in these sources,
and would in particular pinpoint the accelerators of cosmic rays. Good
candidates for high-energy neutrino emission are sources of high energy
gamma-rays, which have e.g.\ been observed with the H.E.S.S.\ atmospheric
Cherenkov telescope {\cite{HESS}}. In spite of the recent significant progress in
the field of high-energy neutrino astronomy, the observation of cosmic neutrinos
from such sources is likely to require cubic-kilometre scale neutrino
telescopes, such as IceCube {\cite{IceCube}} and KM3NeT.
 
In this article two possible configurations of the KM3NeT neutrino telescope are
studied to estimate the sensitivity to generic point sources (assuming a
neutrino flux proportional to $E^{-2}$, where $E$ is the neutrino energy), and
to selected H.E.S.S.\ sources (estimating the neutrino fluxes from the observed
gamma spectra). The full chain from simulation to reconstruction has been taken
into account in this study.

\section{KM3NeT detector configurations}

Several detector configurations, including different geometrical layouts and
optical sensors, are being considered in the KM3NeT Design Study. The main
parameter describing the physics sensitivity of a neutrino telescope is the
neutrino effective area. Due to the energy dependence of neutrino interactions
and the range of high-energy muons in the detector medium (i.e.\ sea water), the
effective area is a function of the neutrino energy and also of the zenith angle
of the direction of observation.

Effective areas for different possible KM3NeT detector configurations have been
obtained from a full simulation chain, including: simulation of neutrino-nucleon
interactions in the vicinity of the detector (only muon-neutrino charged-current
reactions considered); muon propagation and simulation of Cherenkov photons from
muons and secondary charged particles; light propagation in the sea water;
photon detection by the photo multipliers in the optical modules; reconstruction
of the muon tracks from the recorded hits. 

In this paper, two different KM3NeT example configurations are presented. The
first one is based on already used and well known technology, following the
ANTARES design. The second one is a prospective telescope with increased
dimensions and a new type of optical modules housing 21 $3"$ PMTs each,
distributed over the lower hemispheres of the optical modules \cite{multiPMT}. 
In the following, these configurations are labelled ``1'' and ``2''.

Configuration~1 consists of 127 detector strings arranged in a homogeneous
hexagon, each with 25 storeys with ANTARES-type optical modules
($3\;\times\;10"$ PMTs). The distance between lines is 100\,m and between
storeys 15\,m. The simulation was done with the code NESSY, described in
{\cite{Carr}}.

For configuration~2 \cite{Kuch}, a geometrical layout with 225 ($15\times15$)
strings arranged in a cuboid grid is assumed. The distance between the strings
is 95\,m, each string having 36 storeys, separated by 16.5\,m, with one
multi-PMT optical module each.

\begin{figure*}[th]
\begin{center}
\includegraphics [width=0.6\textwidth]{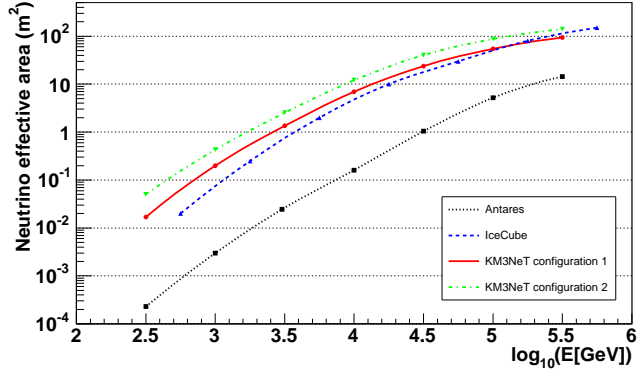}
\end{center}
\caption{
 Neutrino effective areas for the considered KM3NeT configurations. The ANTARES
 and IceCube \protect\cite{Desiati:2006qc} effective areas are shown for
 comparison.
\label{fig3}}
\end{figure*}

The neutrino effective areas obtained for both configurations are compared to
the ANTARES and IceCube effective areas in figure~\ref{fig3}. The effective
areas are calculated applying appropriate selection and reconstruction
requirements and are given as functions of $E$. Note that the program chains
used to study both configurations are completely independent of each other. As
expected, the effective area for configuration~2 is about 2--3 times larger than
for configuration~1, as its instrumented volume and total photocathode area are
larger. In terms of the latter, both configurations exceed IceCube by factors of
about 2 and 3, respectively, which explains the increased effective areas.

The optimal energy interval for the search for neutrino point sources covers 3
decades from 1\,TeV to 1\,PeV. The event rate $R(\Omega)$ from a neutrino source
with celestial coordinates $\Omega$ can be calculated as
\begin{equation}
  R(\Omega) \approx \int \Phi(E,\Omega) A_{\nu} (E,\theta(t,\Omega))\,dE\,dt\,,
  \label{eq-evrat}
\end{equation}
where $\Phi(E,\Omega)$ is a neutrino flux emitted by the source,
$A_{\nu}(E,\theta)$ is the neutrino effective area for the considered telescope
configuration and $\theta(t,\Omega)$ is the zenith angle of the source in
topocentric coordinates at time $t$ (with a period of one sidereal day). The
location of the ANTARES telescope was used to define the local coordinate
system.

\section{Neutrinos from point sources}

The sensitivity of a neutrino telescope to point sources strongly depends on the
energy dependence of the neutrino flux. The current best limit on the neutrino
emission from point sources comes from the AMANDA-II neutrino telescope
{\cite{AMANDA_lim}} and has been obtained assuming a neutrino flux of the form
$\Phi_{\nu_\mu}^0\cdot(E/1\,\text{TeV})^{-2}$. The average limit at 90\%\,C.L.\
for Northern hemisphere sources, from 1001 days of data (2000--2004), is
$\Phi_{\nu_\mu}^0=5.5\times10^{-11}\,\text{TeV}^{-1}\,\text{cm}^{-2}\,
\text{s}^{-1}$. The ANTARES neutrino telescope will reach a similar level of
sensitivity for Southern hemisphere sources with about 1 year of data taking
(assuming a data taking efficiency of 100\%). The higher sensitivity of a
Mediterranean neutrino telescope to point sources is due to the better angular
resolution of the reconstructed muons.

For this study we have calculated expected muon event rates from the point sources in
KM3NeT for two different cases: a) for a generic point source with an
$E^{-2}$ neutrino flux and b) for selected H.E.S.S.\ sources. Although all
of the more than 30 H.E.S.S.\ sources will be in the field of view of the KM3NeT
neutrino telescope, many of them are expected to be too weak for detection.

The energy distribution of H.E.S.S. sources follows a power law $E^{-\Gamma}$,
with a spectral index $\Gamma$ close to 2. Some sources show the indication of
an exponential cut-off in the range of 1--10\,PeV. Under certain assumptions,
the measured gamma flux can be related to the expected neutrino flux from these
sources; for example, in {\cite{Kelner:2006tc}} the following assumptions were
used:
\begin{itemize}
\item
High-energy gamma rays are produced in proton-proton ($pp$) interactions from
hadron (mainly $\pi^0$) decays; due to the charge invariance of the strong
interaction, $\pi^+$, $\pi^-$ and $\pi^0$ are produced in equal abundance and
with similar energy spectra. Contributions from $p\gamma$ interactions are
neglected.
\item
There is no significant absorption of $\gamma$ radiation in the sources, and
energy losses of charged particles can be neglected; charged pions decay before
interaction.
\item
Due to the neutrino oscillations the number of muon neutrinos $\nu_\mu$
(including anti-neutrinos) from the source is equal to the number of gammas
produced in $\pi^0$ decays.
\end{itemize}

The neutrino flux derived from the H.E.S.S.\ measurements can be
parametrised in the form
\begin{equation}
\Phi(E) = k_{\nu}\left(\frac{E}{1\,\text{TeV}}\right)^{-\alpha}
          \exp\left(-\sqrt{\frac{E}{\epsilon}}\right)\;,
\end{equation}
where $k_{\nu}$ is the flux normalisation factor, $\alpha$ denotes the spectral
index and $\epsilon$ the cut-off energy of the source. These parameters were
calculated in \cite{Kappes}.

\section{Event rates}

Event rates from a generic point source as well as for the selected H.E.S.S.\
sources were calculated according to eq.~(\ref{eq-evrat}). For generic $E^{-2}$
sources (averaged for sources with declinations between $-40^\circ$ and
$-20^\circ$), flux limits as low as $\Phi_{\nu_\mu}^0=7.7\times10^{-12}\,\text{TeV}^{-1}
\,\text{cm}^{-2}\,\text{s}^{-1}$ ($\Phi_{\nu_\mu}^0=2.4\times10^{-12}\,\text{TeV}^{-1}
\,\text{cm}^{-2}\,\text{s}^{-1}$) could be achieved with 1 year of data from 
KM3NeT configuration~1 (configuration~2), indicating an increase in sensitivity
of a factor of up to 20 in comparison to current AMANDA limits.

\begin{table*}[th]
\begin{center}
\caption{
Parameters of selected H.E.S.S.\ sources. The source diameter (in degrees) is denoted
by $\delta$, $\kappa_\nu$ is given in units of
$10^{-11}\,\text{TeV}^{-1}\,\text{cm}^{-2}\,\text{s}^{-1}$ and $\epsilon$ in
TeV.\strut}
\begin{tabular}{|r|c|c|ccc|cc|}
\hline
~ & ~ & ~ & \multicolumn{3}{c|}{with cut-off} & \multicolumn{2}{c|}{no cut-off} \\
~ & Name & $\delta(^\circ)$ & $\kappa_{\nu}$ & $\alpha$ & $\epsilon$ 
         & $\kappa_{\nu}$ & $\alpha$ \\ \hline
1 & Vela X    & 0.80 & 11.75 & 0.98 & 0.84 & 4.52 & 2.09 \\ 
2 & RXJ1713.7 & 1.30 & 15.52 & 1.72 & 1.35 & 5.65 & 2.26 \\ 
3 & RXJ0852.0 & 2.00 & 16.76 & 1.78 & 1.19 & 6.25 & 2.29 \\ \hline
\end{tabular}
\label{table1}
\end{center}
\end{table*}
\begin{table*}[th]
\begin{center}
\vspace*{-4.mm}

\caption{
Event numbers from selected H.E.S.S.\ sources for 5 years of data taking. For
each column, the first number (second number in parentheses) indicates the
predicted number of events with (without) taking into account the energy
cut-off; the numbers after the slash indicate the expected backgrounds from
atmospheric neutrinos.
}
\begin{tabular}{|c|cc|cc|}
\hline
~ & \multicolumn{2}{c|}{configuration 2 } & \multicolumn{2}{c|}{ configuration 1 } \\
~ & $\tau_{1}$ / bgr & $\tau_{2}$ / bgr & $\tau'_{1}$ / bgr & $\tau'_{2}$ / bgr \\ 
\hline

1 &   10.0 (16.0) / 13.0 & 23.6 (34.8) / \cA34.0  & 2.1 (2.9) / \cA1.9  & 5.0  (6.9) / \cA4.2 \\
2 & \cA6.4 (11.2) / 23.3 & 15.8 (25.2) / \cA61.0  & 1.4 (2.2) / \cA8.1  & 3.4  (5.1) /   17.3 \\
3 & \cA6.4 (12.9) / 59.0 & 15.8 (29.2) /   154.5  & 1.4 (2.5) /   19.6  & 3.5  (6.1) /   43.3 \\ \hline
\end{tabular}
\label{Table2}
\end{center}
\end{table*}

The KM3NeT sensitivity to the 3 selected H.E.S.S.\ sources summarised in
table~\ref{table1} has been evaluated. All these are Galactic sources with an
extended emission region, which is larger than the KM3NeT angular resolution
(about $0.2^\circ$ above 10\,TeV). For these sources the assumptions made above
for the neutrino flux calculations are expected to be valid.

The resulting event numbers, assuming an observation time of 5\,years, are given
in table~\ref{Table2}. Calculations were done in the interval
1\,TeV\;--\;1\,PeV, for the two cases of assuming a cut-off on the neutrino
energy spectrum (as indicated in table~\ref{table1}), or not. The left-hand
columns ($\tau_1$ and $\tau'_1$) correspond to calculations using the effective
area including full reconstruction of muon events. Since the muon reconstruction
algorithm is not yet fully optimised for the KM3NeT configurations used and by
construction suppresses down-going muon tracks, the event numbers were also
calculated for a simple selection criterion requiring photon signals from the
muon in at least six optical modules; this is an optimistic but reasonable
estimate. The resulting event numbers are shown in the right-hand columns
($\tau_2$ and $\tau'_2$). The background of atmospheric neutrino event numbers
calculated using the Bartol parameterisation \cite{Bartol} is also given.

\section{Conclusions}

Sensitivity to point sources of the future Mediterranean KM3NeT neutrino
telescope have been considered for two different configurations. Neutrino
effective areas for these configurations have been obtained using a full
simulation and reconstruction chain. In the considered energy interval (about
0.1--100\,TeV) the effective areas of both configurations are larger than the
IceCube effective area. The study indicates that the KM3NeT telescope can
significantly improve the current limits on generic point sources. The expected
number of events from given sources rises with the instrumented volume and the
total photocathode area assumed. With the larger of the two KM3NeT
configurations studied, the brightest H.E.S.S.\ sources would be detectable
within less than a decade. However, one has to note that for these studies the
atmospheric muon background was neglected, and the energy reconstruction has
been treated as perfect.

\section{Acknowledgements}

This study was supported by the European Commission through the KM3NeT
Design Study, FP6 contract no.~011937. R.~Shanidze and S.~Kuch wish to thank
A.~Kappes for providing the code for the calculation of event rates for the
KM3NeT configuration 2.

\bibliographystyle{unsrt}
\bibliography{libros}

\end{document}